\documentclass[twocolumn]{aastex63}
\usepackage{graphics,graphicx}
\hypersetup{urlcolor=blue}
\usepackage{natbib}
\usepackage[outdir=./]{epstopdf}
\usepackage{textcomp,gensymb}
\usepackage{xfrac}
\usepackage{xcolor}
\usepackage{mathtools}
\usepackage[caption=false]{subfig}
\usepackage{amsmath,amssymb}
\usepackage{epsfig}
\usepackage{lineno}
\usepackage{bigints}

\newcommand{\mps}{m\,s$^{-1}$}

\newcommand\kms{km~s$^{-1}$}

\newcommand{\tess}{\textit{TESS}}
\newcommand{\ktwo}{{\textit K2}}

\newcommand{\gaia}{{\textit{Gaia}}}
\newcommand{\code}{\textcolor{black}{MOLUSC}}
\accepted{June 12, 2021}
\submitjournal{AJ}

\shorttitle{}
\shortauthors{Wood et al.}
\graphicspath{ {./Images/} }

\begin{document}

\title{Characterizing Undetected Stellar Companions with Combined Datasets}

\correspondingauthor{Mackenna Wood}
\email{woodml96@live.unc.edu}

\author[0000-0001-7336-7725]{Mackenna L. Wood}
\affiliation{Department of Physics and Astronomy, The University of North Carolina at Chapel Hill, Chapel Hill, NC 27599, USA} 

\author[0000-0003-3654-1602]{Andrew W. Mann}
\affiliation{Department of Physics and Astronomy, The University of North Carolina at Chapel Hill, Chapel Hill, NC 27599, USA} 

\author[0000-0001-9811-568X]{Adam L. Kraus}
\affiliation{Department of Astronomy, The University of Texas at Austin, Austin, TX 78712, USA}

\begin{abstract}
    Binaries play a critical role in the formation, evolution, and fundamental properties of planets, stars, and stellar associations. Observational studies in these areas often include a mix of observations aimed at detecting or ruling out the presence of stellar companions. Rarely can non-detections rule out all possible binary configurations. Here we present \code, our framework for constraining the range of properties of unseen companions using astrometric, imaging, and velocity information. We showcase the use of \code{} on a number of systems, ruling out stellar false positives in the signals of HIP67522b, and DS Tuc Ab. We also demonstrate how \code{} could be used to predict the number of missing companions in a stellar sample using the ZEIT sample of young planet hosts.
    Although our results are not significant, with a larger sample \code{} could be used to see if close-in planets are less common in young binary systems, as is seen for their older counterparts.\\
\end{abstract}

\section{Introduction}
Binaries and higher-order stellar systems are a common outcome of star formation. Roughly half of all Sun-like stars in the field are in multiple star systems \citep{raghavan_survey_2010, duquennoy_multiplicity_1991}, with the multiplicity fraction rising to near unity for intermediate and high-mass stars, \citep{zinnecker_toward_2007, moe_mind_2017}, and falling to $\simeq$30\% for late-type dwarfs \citep{winters_solar_2019}. The frequency and properties of multi-star systems are also known to vary with metallicity, age, and environment \citep{duchene_stellar_2013}. The ubiquity of multiple star systems makes understanding them critical for a wide range of astrophysics, ranging from star and exoplanet formation to stellar clusters and high-energy astrophysics.

With thousands of transiting exoplanet candidates from NASA's {\it Kepler}, {\it K2}, and {\it TESS} missions, it is not feasible to confirm each planet through radial velocity detection. Instead, most planets are statistically validated by, in large part, ruling out the presence of stellar companions or background stars that could reproduce the transit-like signal \citep[e.g., ][]{fressin_false_2013, morton_false_2016, bryson_identification_2013}. Even if a binary or higher-order multiple cannot explain the transit signal, the presence of binaries can dilute the transit depth and alter the inferred properties of the planet \citep[e.g., ][]{ciardi_understanding_2015, bouma_biases_2018}.
This is especially a problem when comparing planet properties between different populations \cite{rizzuto_zodiacal_2017}; since the binary fraction changes with age and metallicity \citep{duchene_stellar_2013}. Observed differences in the planet populations may be contaminated by differences in the undetected companion population. Uncorrected stellar flux in the same aperture can also impact studies of the planet's transmission spectrum by creating a wavelength-dependent dilution \citep{desert_low_2015}.

In addition to changing the derived parameters and classification of planets, stellar companions also directly impact their formation and evolution. Stellar companions can impede the formation or survival of close-in planets \citep[e.g.,][]{wang_influence_2014, kraus_impact_2016, moe_close_2019}. There is also some evidence that they facilitate the formation or migration of close-in Jovian-mass planets \citep[e.g.,][]{ngo_friends_2016}.
 
Beyond exoplanets, unseen binaries can bias isochronal ages of stellar associations \citep[e.g.,][]{malo_banyan_2014, bell_self-consistent_2015, sullivan_undetected_2021}, change stellar parameters inferred from color-magnitude diagrams and stellar spectra \citep{el-badry_signatures_2018}, and skew estimates of the stellar initial mass function \citep{kroupa_effects_1991}. 
In each of these cases, the bias arises because a companion is present but undetected. While it might be possible to account for this bias by modeling the presence of unseen companions, the large range of possible companions and data heterogeneity can make this impractical. 

Even with extensive spectroscopic and imaging follow-up, it is rarely possible to completely rule-out the presence of a binary companion for any given star. While radial velocity and high-resolution imaging are often complementary in terms of their coverage with semi-major axis, the former is insensitive to stars with face-on orbits and the latter to low-mass companions and those directly behind the star at the time of imaging. Even binaries that are easier to detect based on their period and mass ratio alone may be missed due to the timing of the follow-up. These situations are sufficiently rare as to have a small impact on statistical validation of individual exoplanets \citep{morton_low_2011} or when deriving stellar parameters for a specific target. However, as the sample of stars increases, the number of possible missed companions grows, and assuming no such companions are present will systematically bias the result. Instead, many studies may include a measure of their completeness as part of a comparison to the population of field stars \citep[e.g., ][]{wang_influence_2014-1, wang_influence_2015}. This approach, however, requires assumptions about the underlying binary population and the relation between the observed and true properties (e.g., projected separation at a given epoch versus semi-major axis).

Recent approaches, such as those using Monte Carlo methods and rejection sampling to model possible companions from multiple datasets, can provide strong constraints on stellar and planetary companions \cite[e.g.][]{boehle_combining_2019, hurt_decade_2021, lagrange_unveiling_2020, blunt_orbits_2017, hinkley__2013}. \citet{boehle_combining_2019} used Monte Carlo modeling of planetary companions, and a combination of high resolution imaging and radial velocity measurements of several nearby star systems to study possible planetary configurations in those systems. We build upon these efforts but focus primarily on stellar companions. We also include additional constraints from \gaia{} imaging and astrometry, which have been shown to be effective in identifying and constraining binaries \citep{ziegler_measuring_2018, brandeker_contrast_2019, kraus_impact_2016}.

Here we present \code{} \footnote{https://github.com/woodml/MOLUSC} (\underline{M}ulti-\underline{O}bservational \underline{L}imits on \underline{U}nseen \underline{S}tellar \underline{C}ompanions), a open-source framework and program that provides full posteriors on possible companions for a target based on a suite of possible input data. \code{} generates a grid of binaries under a physically-motivated, and user-adjustable distribution, which it then compares to \gaia{} astrometry and imaging, and user-provided contrast curves and radial velocities. The output is a full posterior on potential surviving stars (or giant exoplanets). We use this to investigate the number of unseen binaries in a sample of some of the known, young planets from {\it K2}.

In Section~\ref{sec:code} we present the basic framework for the code, including the default assumptions, user-adjustable parameters, and possible inputs. We apply this to a number of test cases in Section~\ref{sec:tests}. In Section~\ref{sec:ZEIT}, we apply this to the sample of known young planets and demonstrate that there are $\simeq1.6$ undetected companions around the twelve planet-hosts, in addition to one already identified. This is consistent with the population of older planet hosts, and the range of possible missed companions is not enough to significantly alter the planet radius distribution. We conclude in Section~\ref{sec:conclusions} with a summary and a brief discussion of the applications for \code{} beyond the intended application for exoplanet vetting and binary statistics. 

\section{Methodology}\label{sec:code}

Our aim is to generate realistic binary probabilities when no companion is detected. To this end we use a Monte Carlo simulation of possible companions to the primary star. We compare the generated orbital properties to observational constraints from High Resolution Imaging, \gaia{} Imaging, RV measurements, and the fit quality of the \gaia{} astrometry and use rejection sampling to determine posteriors and detection limits on the system. 

In order to densely sample the parameter space, especially at low masses and periods, at least 1 million companions should be generated. For each of the test cases described in Section \ref{sec:tests} we generate between 1 and 10 million companions. Sections of the code are parallelized for efficiency, so the runtime will vary significantly with number of cores used, as well as with the types of analysis chosen.

\subsection{Companion Generation} \label{sec:generation}
To fully describe a binary system you need a description of both the component orbits, and the photometric parameters. 

\begin{figure*}[t]
    \centering
    \includegraphics[width=\linewidth]{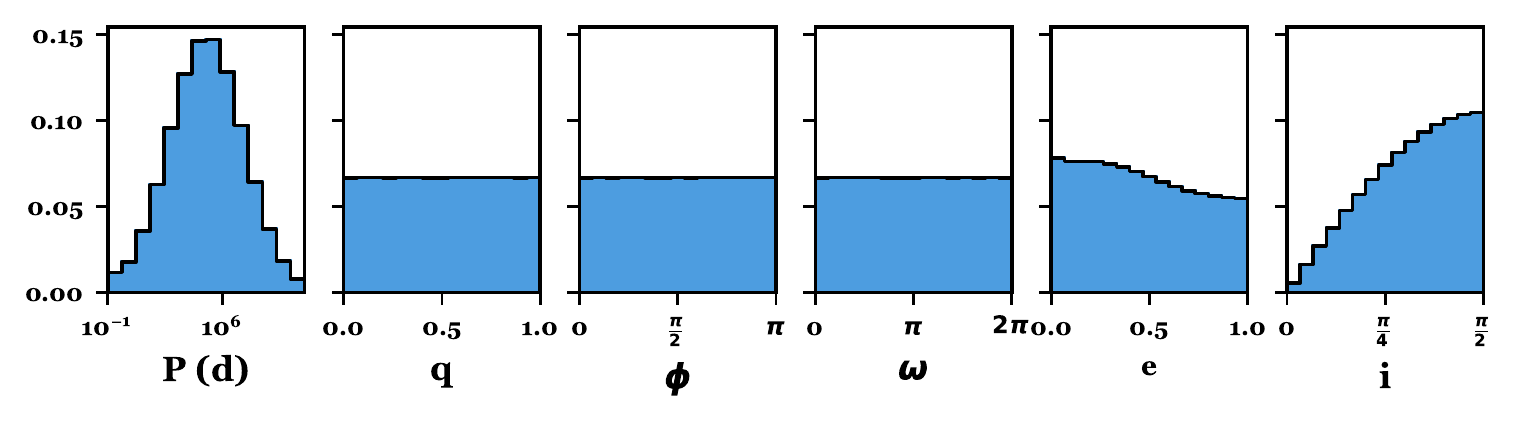}
    \caption{Example generated prior distributions for the six generated orbital parameters (described in Table \ref{table:notation} and Sec \ref{sec:generation}) for a sample of 5 million companions. Period is distributed log-normally in days with a default mean and standard  deviation of $5.03$ and $2.28$. Mass ratio is generated using a power-law distribution, with a default exponent of $0.0$, creating a uniform distribution. Pericenter phase, and argument of periapsis are distributed uniformly within appropriate bounds. Eccentricity follows a two-part distribution dependent on period, discussed in Sec \ref{sec:eccentricity}. Inclination is generated uniformly as $\cos(i)$. We note that due to the effects of random number generation the uniform distributions are not perfectly uniform.}
    \label{fig:all_params}
\end{figure*}

An orbit is uniquely described using 6 parameters. We elect to use period ($P$), inclination ($i$), mass ratio ($q$), eccentricity ($e$), argument of periapsis ($\omega$) and pericenter phase ($\varphi$) to describe the orbits of the simulated companions. Our choice of parameters differs from that used by other authors, for example, by excluding longitude of ascending node, as it affects neither the RV calculation nor the projected separation calculation (see e.g. the discussion of RV calculation given by \citet{perryman_exoplanet_2011}).  
Each of these generated parameters is constant for a given system. The primary mass, which is constant for all generated systems, is used to calculate additional system parameters.
Parameters that change over time, such as the radial velocity or projected separation, are calculated from the generated parameters as needed. Parameters and priors are summarized in Table \ref{table:notation}. The parameter distributions are summarized in Figure \ref{fig:all_params}, and described in detail below.

\code{} randomly generates each orbital parameter following realistic parameter distributions, described in Sections \ref{sec:period}-\ref{sec:other}, for the number of hypothetical binaries specified by the user. The user can also choose to fix parameters or limit them by a minimum or maximum value, depending on the situation. In this work we limit the parameter distributions to values for stellar-mass companions, and do not accurately model orbital parameter distributions for planetary companions, as has been done previously by other authors \cite[e.g.][]{boehle_combining_2019}. However, the code framework itself could easily support this, and is trivially adjustable to set an appropriate parameter distribution for planetary-mass companions.

We use the additional parameters system age, primary mass and stellar jitter to describe the photometric properties of the system. Each of these is provided by the user, and used to model properties such as luminosity and activity.

\begin{table}[ht!]
    \centering
    \begin{tabular}{|c|c|c|}
     \hline
     Notation & Meaning & Distribution\\ 
     \hline
    P & period & Log Normal\\
    q & mass ratio, $M_c/M_p$ & Uniform\\
    i & inclination & Uniform in cos(i) \\
    e & eccentricity & See \ref{sec:eccentricity}\\
    $\omega$ & argument of periapsis & Uniform\\
    $\varphi$ & pericenter phase & Uniform\\
    \hline
    $M_p$ & Primary Mass & User-provided \\
    Age & Stellar Age & User-provided\\
    \hline
    a & Semi-major axis & Calculated \\
    $\rho(t)$ & Projected separation  & Calculated \\
    $RV(t)$ & Radial velocity & Calculated \\
    $\Delta M$ & Contrast (arbitrary filter) & Calculated \\
    
     \hline
    \end{tabular}
    \caption{Explored stellar parameters. The rightmost column describes either the distribution used, if the parameter is generated randomly, or otherwise, the method of obtaining it. The first 6 are the generated orbital parameters, discussed in detail in Section \ref{sec:generation}. The next two are stellar parameters, provided by the user for the target star. The last four are derived parameters, which we calculate as needed using the orbital or stellar parameters and stellar evolutionary models, discussed throughout Section \ref{sec:data}}.
    \label{table:notation}
\end{table}

\subsubsection{Period} \label{sec:period}
The period distribution depends strongly on primary mass \citep{duchene_stellar_2013, moe_mind_2017}. A log-normal distribution is often found for solar-mass \citep[e.g.][]{heacox_logarithms_1996, raghavan_survey_2010} and low-mass stars \citep{fischer_multiplicity_1992, janson_astralux_2012, janson_astralux_2014}, with the mean and standard deviation increasing with increasing mass \citep{duchene_stellar_2013}. Higher mass stars have been suggested to follow a more bimodal distribution, with peaks at $\log(P(d))=1, 3.5$ \citep{moe_mind_2017}. 
However, the period distribution of O and B stars is poorly defined, because high mass stars often occur in cluster environments, where it is difficult to detect wide binaries \citep{sana_southern_2014}. Measuring the period of wide binaries is also difficult  (c.f. Figure 1 of \citet{moe_mind_2017}), and few surveys have large samples of wide, high mass binaries \citep{sana_southern_2014}.
Given this, and that the models we used do not extend to high mass stars, we decide to use the distribution for solar-type stars from \citet{raghavan_survey_2010} as the default for all simulations. This distribution should be accurate for F, G, and K dwarfs in the Solar neighborhood \citep{raghavan_survey_2010, duchene_stellar_2013}. The user has the choice to use a different log normal distribution for other stellar types by providing the mean and standard deviation.

Regardless of the distribution used, we apply a lower limit to the period distribution of 0.1 days. This avoids numerical difficulties in calculating radial velocities at extremely short periods. The 0.1-day limit was selected because at that period any companion is likely to be interacting significantly with the primary, hence the assumptions made in this code will no longer be valid. 

\subsubsection{Mass Ratio}

The binary mass ratio distribution is commonly characterized using a piecewise function with one function for values of $q$ between $0$ and $0.95$ and a second for values between $0.95$ and $1$. This accounts for the known significant increase in frequency of nearly equal-mass binaries at $q\gtrsim 0.95$ \citep{lucy_significance_1979, raghavan_survey_2010, kounkel_close_2019, el-badry_discovery_2019}, known as the twin excess.

The distribution below the twin excess depends on primary mass \citep{duchene_stellar_2013, el-badry_discovery_2019, moe_close_2019} and  can be described using either a power-law distribution \citep[e.g.,][]{moe_close_2019}, or a uniform distribution \citep[e.g.,][]{raghavan_survey_2010}. \citet{raghavan_survey_2010} and \citet{kounkel_close_2019} found that a uniform distribution applies for a broad range of masses near solar mass.
We choose to use a uniform distribution (a power law distribution with exponent zero) as the default, but the user can choose to use a different power law distribution. Per the results from \citet{el-badry_discovery_2019}, \citet{kounkel_close_2019} and \citet{moe_mind_2017}, we believe this default uniform distribution should be accurate for systems with primary mass $0.1 \lesssim M_{1} \lesssim 1.5 M_{\odot}$, and suggest using a different power law exponent for masses outside that range.

We do not include the twin excess in the mass ratio distribution for two reasons: as shown in \cite{moe_mind_2017} and \cite{el-badry_discovery_2019}, the twin excess is not present at all periods or primary masses; and when it is present, the amount of excess is poorly constrained. 
Therefore, we use a single mass ratio distribution over the full  range of $ 0 < q < 1$. Since equal-mass companions are generally the easiest to detect, excluding the twin excess has only a small impact on the posteriors.

We do not apply a lower limit to the mass of generated companions, allowing brown dwarf and planet companions to be generated along with stellar companions. Since all companions are generated using the stellar distributions, planet-like companions will not necessarily follow a physical  distribution of planetary mass companions. Thus, the default settings for this code are not suitable to determine the possible parameter space of missing planet companions, but only for stellar companions. It can still be used to rule out stellar false positives in data with planet detections, as demonstrated in Section \ref{sec:tests}.

\subsubsection{Eccentricity}\label{sec:eccentricity}
At the shortest periods, all binaries have nearly circular orbits \citep[e.g.,][]{price-whelan_close_2020}, due to tidal circularization \citep{zahn_tidal_1977}. At large periods, the distribution is uniform \citep[e.g.,][]{raghavan_survey_2010}. However, in the middle regime the statistical upper extent of eccentricity increases with period \citep{duchene_stellar_2013, price-whelan_close_2020, murphy_finding_2018}.

\begin{figure*}[t]
    \begin{center}
    \includegraphics[width=\linewidth]{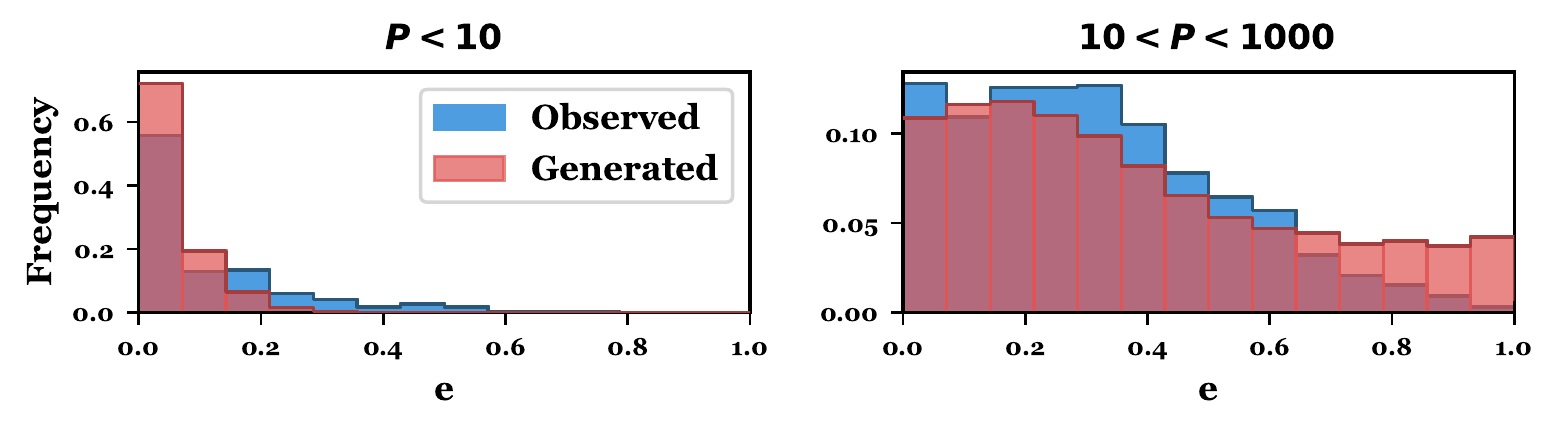}
    \caption{A comparison of the observed eccentricity distribution from \citet{price-whelan_close_2020} and \citet{murphy_finding_2018} and the model distribution described in Sec \ref{sec:eccentricity}. Blue: The observed distribution. Red: A sample generated distribution. Bins are the same in all histograms. Below $10$ days nearly all systems are circular, with increasing numbers of high eccentricity systems at increasing periods. Our generated distribution has fewer low period - high eccentricity systems, and more high eccentricity systems at long periods. However, these differences may be caused by observational biases (see text), and overall, the generated distribution closely resembles the observed distribution for close binaries.}
    \label{fig:eccentricity}
    \end{center}
\end{figure*}

To explore distributions of eccentricity at $P < 1000$ days we used a sample of 1,347 binary stars compiled from \cite{murphy_finding_2018} and \cite{price-whelan_close_2020}. These samples cover a range of periods between $0$ and $10^4$ days, and contain a large number of randomly sampled systems, with stellar types including FGK dwarfs and A stars. We show the eccentricity distribution from this combined sample in blue in Figure \ref{fig:eccentricity}.

We choose to model the eccentricity in this period range as a normal distribution, where the mean $\mu_e$, and standard deviation $\sigma_e$, are functions of $P$, as shown in the equations below.
\begin{linenomath*}
\begin{gather}
    \mu_e = a\log_{10}(P) + b \\
    \sigma_e = c\log_{10}(P) + d,
\end{gather}
\end{linenomath*}
Here, $a$, $b$, $c$, and $d$ are fit parameters, that we determine using a maximum likelihood comparison on the data sample described above.
We then use the calculated values of $a,b,c,$ and $d$ to define Gaussian distributions from which we draw eccentricity for companions with $P < 1000 $ days.

\begin{table}[h!]
    \centering
    \begin{tabular}{|c|c|}
        \hline
        Parameter & Value \\
        \hline
        a & 0.148 \\
        b & 0.001 \\
        c & 0.042 \\
        d & 0.128 \\
        \hline
    \end{tabular}
    \caption{The values of the eccentricity fit parameters resulting from a maximum likelihood fit on a sample of close binaries from \citet{price-whelan_close_2020} and \citet{murphy_finding_2018}.}
    \label{table:eccentricity_fit}
\end{table}

At periods longer than $1000$ days we draw eccentricities from a uniform distribution between 0 and 1. This aligns with the assessment in \cite{duchene_stellar_2013} that eccentricity is uniform at long periods. We chose $1000$ days as the transition point to create a smooth transition between the two regions. We also apply a limit on the maximum value of eccentricity, since eccentricities of exactly $1.0$ produce unphysical radial velocities (and are unlikely to remain bound). The maximum eccentricity is $0.9999$. 

A typical distribution of the eccentricities and periods of generated binaries is shown in red in Figure \ref{fig:eccentricity}. This qualitatively matches the observed distribution of eccentricities, as shown in blue in Figure \ref{fig:eccentricity}.

There are some small differences between our generated sample and the empirical results. One is that our parameterization has a low probability of creating binaries with $P < 10$ days and $e > 0.2$. \cite{price-whelan_close_2020} show several such binaries (see left column of Figure~\ref{fig:eccentricity}). However, \citet{price-whelan_close_2020} note that of their systems with $P \lesssim 5 $ days and $ e \gtrsim 0.4$ are likely caused by systematic uncertainties, poor sampling, and larger random uncertainties on the eccentricities in this regime. There is also a difference in the number of high-eccentricity systems compared to \citet{price-whelan_close_2020} and  other findings in the literature \citep[e.g.,][]{stepinski_orbital_2001, murphy_finding_2018}, which find fewer high-eccentricity, long-period systems. We elect not to include a decrease in likelihood for high eccentricity systems at long periods. Because the difference is mostly at wide separations, these are also the systems that are generally easy to detect, so this had a small impact on the results. To test this we generate 4 sets of 1 million companions, all with $P > 5,000$ days. In two of the sets no companions with $e > 0.9$ are generated. We run High-Resolution Imaging and RV tests (see Sections \ref{sec:AO} and \ref{sec:RV} for details) on these sets, and find that the presence or absence of high-eccentricity, long-period binaries did not significantly change the posterior distributions.

\subsubsection{Other parameters}\label{sec:other}
Argument of periapsis ($\omega$) is drawn from a uniform distribution between 0 and $\pi$. Argument of periapsis measures the angle between the line of sight and the system pericenter. We define $\omega$ so that it is $0$ along the line of sight. 
The pericenter phase ($\varphi$) is drawn from a uniform distribution between 0 and $2\pi$. We define $\varphi$ as the angle between the companion and the orbit pericenter at reference time $T_0$. 

Inclination $i$ is not uniformly distributed on the sky. For a random distribution of orientations, a face-on view is less likely to occur than an edge-on view, as there are more possible configurations that could result in an edge on view than a face-on one. The probability distribution of inclination is thus proportional to $\sin i$ \citep{heintz_statistical_1969}. However, a simpler way to achieve the same distribution is to generate along $\cos i$ following a uniform distribution from 0 to 1, which is the method we utilize here.

\subsection{Input Data}\label{sec:data}

\begin{figure}[h!]
    \begin{center}
    \includegraphics[width=\linewidth]{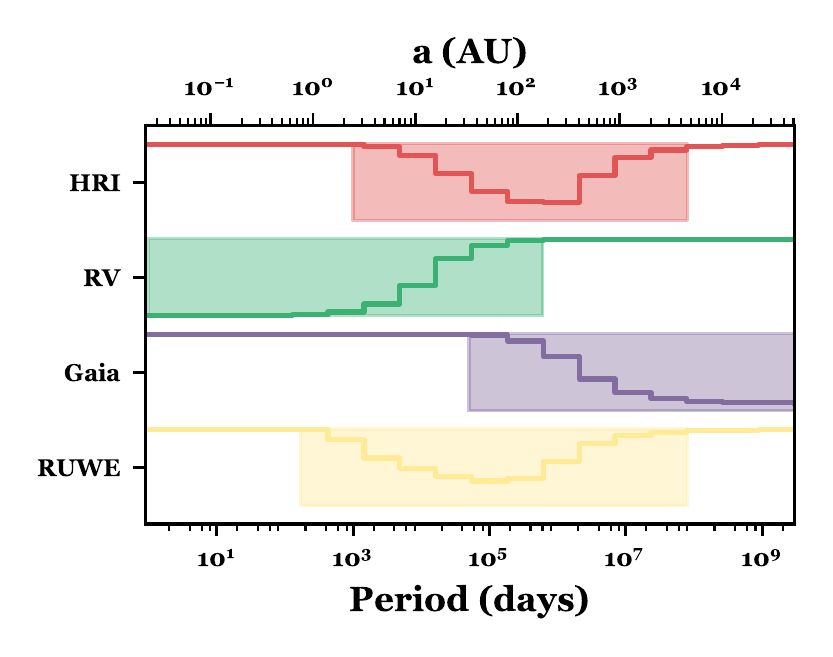}
    \caption{The effective period ranges of each of the tests, for the star HIP67522 (see Section \ref{sec:hip67522}). The HRI test is shown on top in red, RV second from top in green, \gaia{} imaging in purple second from bottom, and \gaia{} astrometry (RUWE) in yellow on the bottom. The lines show the surviving fraction of generated binaries in period bins, where bins are the same for each line. The boxes show the effective range of each test. The top and bottom of each box show a survivor fraction of $100\%$ and $0\%$, respectively. The effective ranges vary with distance, and will shift towards longer periods in stars with larger distances.}
    \label{fig:comparison}
    \end{center}
\end{figure}

A key feature of \code{} is the ability to combine multiple types of data and obtain the strongest limits on the system. \code{} can use user-provided follow up data from High Resolution Imaging (HRI) and RV measurements, and can use imaging and astrometric fit constraints from \gaia{}. These four types of data are complementary, combining to create strong limits on the systems. 
RV analysis works best on close-in companions, viewed from an edge-on angle. In contrast, high resolution imaging works best on intermediate period systems, viewed face-on. For most systems the periods best explored by RV fall just below those explored by HRI. \gaia{} imaging constraints overlap with HRI constraints, and continue past them, while RUWE constraints cover a regime overlapping with all three of the other tests. 
We demonstrate this in Figure \ref{fig:comparison}, showing the effective period ranges of each test, for the star HIP67522 (See details in Section \ref{sec:hip67522}). This star is magnitude 9.6 in \gaia{} G, at a distance of $128$ pc. The effective ranges shown here will be different for stars at different distances, since the HRI and \gaia{} tests depend on projected separation, and will shift towards longer periods at larger distances.  

\subsubsection{High Resolution Imaging} \label{sec:AO}

Contrast curves, constructed from high resolution imaging (HRI), such as adaptive optics imaging, speckle data, or interferometry, can be used to place limits on magnitude and projected separation and are routinely taken as part of exoplanet or similar follow-up programs \citep[e.g.,][]{Furlan_imaging_2017}. The user can provide such curves, derived from any method or telescope, as lists of contrast detection limits as a function of projected separation.
We use these contrast curves to place constraints on possible companions by using evolutionary models to relate companion magnitude and mass, and using the orbital parameters to calculate projected separation.

To determine the magnitudes of the primary and simulated companions, we use the models from \citet{baraffe_new_2015} in the relevant filter.  We have implemented testing in the 2MASS $J$, $H$ and $K$ filters, \gaia{} $G$, $Bp$ and $Rp$, and CFHT $R$, and $I$ filters. The \citet{baraffe_new_2015} models span ages from $0.0005 $ Gyr to $10 $ Gyr and masses from $0.01 M_\odot$ to $ 1.4 M_\odot$. If the primary star is outside of these age or mass ranges the user can adjust the code to use a model appropriate for the target.
The mass of the primary, given by the user, and the generated masses of the companions are plugged into the model, using linear interpolation between grid values when necessary, to calculate the magnitude. Companions with masses below the range available in the model are assigned contrast values of infinity, and hence will always be considered undetectable by HRI.

\begin{figure}[ht!]
    \begin{center}
    \includegraphics[width=\linewidth]{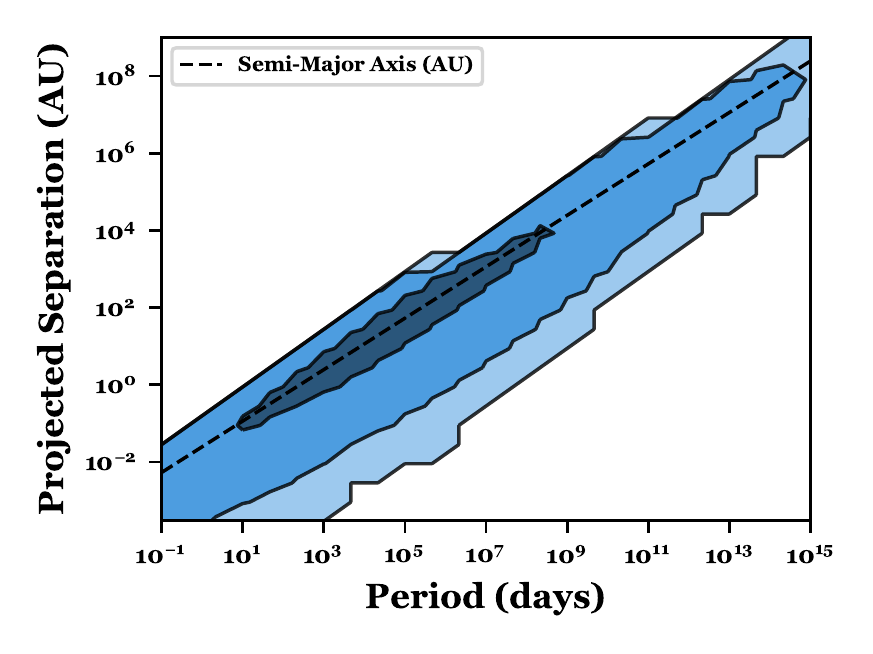}
    \caption{A contour plot showing projected separation vs. period for a sample of 5 million simulated stars. All orbital parameters are generated as described in Section \ref{sec:generation}. The dashed line indicates the semi-major axis for a given period, assuming an equal mass binary with a $1M_{\odot}$ primary. Contours are at the 39.8, 86 and 98th percentiles of density. Projected separation is often significantly lower than the true semi-major axis, meaning that companions that would be seen at the distance of their semi-major axis can be obscured in the higher detection limits closer to the primary. Due to non-zero eccentricity, binaries can also be at separations larger than their semi-major axis, making them easier to identify at specific epochs.}
    \label{fig:projected_sep}
    \end{center}
\end{figure}

To complete the comparison of the companions to the contrast curve, we calculate the projected separation of each simulated system at the time of observation(s). By generating a full orbital solution for each companion, and then using that to calculate the true projected separation, rather than the semi-major axis, the posterior probabilities produced by \code{} are more accurate than those produced by assuming that the separation is the semi-major axis. As we show in Figure \ref{fig:projected_sep}, detection limits which treat the projected separation as the semi-major axis will be overly optimistic or introduce systematic biases towards eccentric systems. For an equal mass binary with $ P=5000$ days and $a = 7.95 $ AU, $18 \%$ of the generated companions had $\rho > 1.3a$ and $30\%$ had $\rho < 0.7a$.

We calculate the projected separation using Equation~\ref{eqn:projected_sep}, where $f(t)$ is the true anomaly, as defined by \citet{perryman_exoplanet_2011}.
\begin{equation}\label{eqn:projected_sep}
    \rho = a \frac{1-e^2}{1+e\cos(f(t))} \sqrt{\sin^2{(f(t)+\omega)}+\cos^2{(f(t)+\omega)}\cos^2(i)}
\end{equation}

Our formulation of projected separation differs from that used by some other authors \citep[e.g.][]{kane_maximum_2018}. This difference is likely due to differences in the definition of $\omega$. We define $\omega$ so that it is zero along the line of sight. 
The time, $t$, at which projected separation is calculated can be included in the user-provided contrast file. Doing so increases the power of simultaneous constraints for systems with long enough RV monitoring to overlap with the imaging constraints, or from additional HRI data.

The next step is to convert the given contrast curve from angular projected separation (e.g. in mas) to physical projected separation (e.g. in AU). We obtain the distance to the target star using \gaia{} by searching an area of 10 arcseconds around the provided primary star coordinates and adopting the closest object as the primary. We use the \gaia{} parallax of the object to calculate distance, and convert the angular separations given in the contrast curve to projected separations in AU.
The contrast, $\Delta M$, between the primary and each companion can then be calculated and directly compared to the provided contrast curve.

We calculate the experimental limit on $\Delta M$ for the projected separation of each system by linearly interpolating between the points of the given contrast curve. If the modeled $\Delta M$ is less than the experimental limit on $\Delta M$, the companion star would have been visible and is rejected. If the modeled contrast is greater than the experimental limit it fails to be rejected. 

In many cases contrast curves are created with recovery probabilities as a function of contrast and separation. This is common for contrast curves derived using injection/recovery tests \citep[e.g.,][]{marois_exoplanet_2010}. In such cases, we reject companions with a probability matching the recovery rate at the companions' projected separation and contrast. We use linear interpolation to fill in between the given points of the contrast curve.

\subsubsection{\gaia{} Imaging} \label{sec:Gaia}

\gaia{} DR2 is estimated to be complete to 18th magnitude across the whole sky and 20th magnitude across most of the sky \citep{arenou_gaia_2018}. The completeness remains near 100\% for binaries with separations greater than 4\arcsec{} down to the limiting magnitude \citep{arenou_gaia_2018}. Therefore, companions with magnitudes brighter than $G=18$, and projected separations greater than 4\arcsec{} are rejected. This hard limit may be overestimating the completeness, since it is possible that a companion is hidden behind a neighboring star, but this effect will be small outside of crowded fields.
We assume that the star does not have a companion visible in \gaia{} DR2. If such a companion does exist the inclusion of \gaia{} imaging input in the analysis will produce incorrect results in the parameter space of the companion. It is left to the user to ensure that there are no comoving companions visible there, as we do not explicitly check nearby star proper motions. Treatment of a star with a resolved companion in \gaia{} is demonstrated in Section \ref{sec:ds_tuc}.

Companions that are closer or fainter than the \gaia{} completeness limit can still be detected by \gaia{} some of the time, and are rejected with rates proportional to their detection probability. 
To quantify the completeness at separations of less than 4\arcsec{} we use the recovery rates found by \citet{ziegler_measuring_2018}. They used Robo-AO to determine the binary recovery rates of \gaia{} DR2 and found that binaries with magnitude contrasts up to 6 are recovered down to 1\arcsec.  These limits reach to separations of slightly over 3\arcsec, complementing the \gaia{} completeness limit at 4\arcsec.
Recovery rates for companions with separations larger than 4\arcsec, and contrasts below the \gaia{} completeness limit are obtained from \citet{brandeker_contrast_2019}. They found recovery rates for \gaia{} out to $12,000$ mas, and down to $\Delta G \simeq 14$.
We use the recovery rates from these papers to determine the likelihood that a companion with a given separation and magnitude would have been detected by \gaia{}, and can be rejected.

\begin{figure}[ht!]
    \centering
    \includegraphics[width=\linewidth]{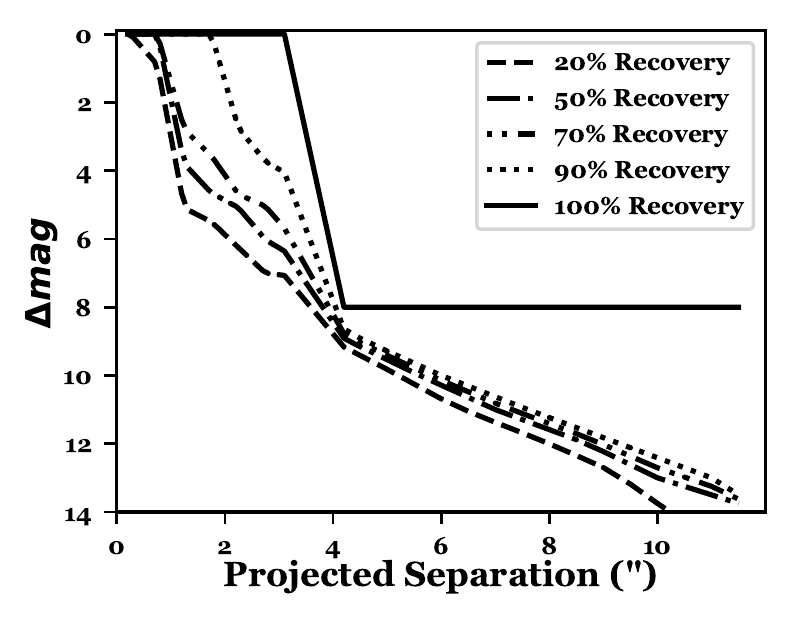}
    \caption{A schematic showing \gaia{} contrast limits for a 10th magnitude star. For projected separations less than 4000 mas, recovery rates from \cite{ziegler_measuring_2018} are used. At separations greater than 4000 mas we use the \gaia{} completeness limit of 18th magnitude, and the recovery rates from \cite{brandeker_contrast_2019}. If the limits from \cite{ziegler_measuring_2018} and \cite{arenou_gaia_2018} are in conflict, as might occur if the primary is dimmer than 12th magnitude, the \gaia{} detection limits are used.}
    \label{fig:gaia_limits}
\end{figure}

The \gaia{} completeness limits and the recovery rates from \citet{ziegler_measuring_2018} and \citet{brandeker_contrast_2019} are combined to create a single contrast curve (shown schematically in Figure \ref{fig:gaia_limits}), which we treat as a HRI contrast curve as described in Section \ref{sec:AO}. For this data, we assigned an epoch of 2016.0, i.e., the typical value for \gaia eDR3 astrometry.

\subsubsection{Radial Velocity} \label{sec:RV}

RV measurements can place strong constraints on nearby companions that complement the data provided by imaging and astrometry. For each simulated binary, we compute the corresponding radial velocity curve, and compare this to the observed RV curve to determine which companions could be present in the system.

To simulate the RV curve, we first determine if a system would be an SB1 or an SB2 by examining the magnitude contrast in G between the components. For pairs with $\Delta G >5$ the flux contribution of the companion is considered negligible, and the system treated as an SB1. Pairs with $\Delta G < 5$ are considered SB2s, and their treatment determined on whether they would be resolved or not based on the provided RV measurements.

We calculate the RVs following the procedure by \citet{perryman_exoplanet_2011}. For SB1 systems we calculate the RV curve of only the primary component, whereas for SB2 systems we calculate RV curves for both primary and companion.

First, we calculate the radial velocity semi-amplitude, $K$, in terms of the generated parameters:
\begin{equation}
    K = \frac{2\pi}{P}\cdot\frac{q}{q+1}\cdot\frac{a\cdot\sin(i)}{\sqrt{1-e^2}}
\end{equation}
Once the semi-amplitude is calculated, we compute the the time-dependent portion of the radial velocity. This depends on the true anomaly, $f(t)$, as described by \citet{perryman_exoplanet_2011}. 
We combine the amplitude and time-dependent components to calculate the radial velocity:
\begin{equation}
    RV(t) = K\cdot\cos(\omega+f(t)) + e\cdot\cos(\omega)
\end{equation}

We divide SB2's into resolved and unresolved groups, based on the velocity difference between the two components at each epoch of observation, and the user-provided spectral resolution of the RV measurements. Systems with larger velocity differences than the spectral resolution would have been resolved SB2s, and are rejected. Composite RV curves for unresolved SB2 systems are calculated by taking a flux-weighted average of the primary and companion RVs.
We then compare the unresolved SB2s and SB1s to the measured RVs as described below.

We assume that the barycentric velocity of the system is unknown and account for it by applying a zero-point shift to the predicted RVs. We perform a least squares test on the predicted and measured RV curves and choose the zero point shift that produces the best fit between them. 

Once the best-fit zero point is applied, we perform a $\chi^2$ test, with degrees of freedom equal to the number of RV measurements minus one, to determine the goodness of fit between the measured and predicted velocities. 

In addition to the modeled RV variation due to a companion, a given star may have "stellar jitter", the extra variation in its RV curve not accounted for by orbit parameters (see e.g. \citet{oshagh_understanding_2017, meunier_activity_2019}). Without accounting for this variation, companions may be rejected due to the stellar noise. In the case of a particularly noisy star, it is possible that no simulated RV curve would fit the noiseless data, and all companions would be erroneously rejected. To correct for this, the user can choose to provide a value of stellar jitter, which is added to the RV error in quadrature during the $\chi^2$ test.
The $\chi^2$ cumulative distribution function probability is then taken as the rejection probability of the system.

The rejection probability alone is not sufficient to determine whether a companion could have been present but unobservable. Small radial velocity variations could be below the sensitivity of the detector, rendering them unobservable. To account for this the user can provide a RV sensitivity floor, specifying the smallest RV amplitude which could have been detected in the data. We reject or keep the simulated SB1 and unresolved SB2 systems depending on both the RV amplitude and their $\chi^2$ rejection probability. Companions with $\text{Amp} < \text{RV Floor}$ are not rejected regardless of rejection probabilities. Those with $\text{Amp} > \text{RV Floor}$ are rejected according to their $\chi^2$ rejection probabilities.

\subsubsection{\gaia{} Astrometry (RUWE)}\label{RUWE}

The precise astrometric measurements of \gaia{} eDR3 can be used as an additional test for binarity. A large excess of astrometric noise in \gaia{} has been shown to be a signature of an unresolved stellar companion  \cite[e.g.][]{ziegler_soar_2019, pearce_orbital_2020, belokurov_unresolved_2020}. This is likely because \gaia{} DR2 and eDR3 reduction treats all stars as single; a companion may both introduce astrometric variation and cause a mismatch between the true point-spread function and the one used in extracting the astrometry. Accordingly, lack of large astrometric noise can be used to constrain the range of possible binary companions. Using a set of stars that are unresolved in \gaia, but with known companions (e.g., from high-resolution adaptive optics) it is possible to calibrate this effect and include it in our model.

The \gaia{} Re-normalized Unit Weight Error (RUWE) is a measure of the astrometric noise, re-normalized to correct for differences in color and magnitude (\citet{lindegren_gaia_2018}, \url{https://www.cosmos.esa.int/web/gaia/public-dpac-documents}). It is a reduced $\chi^2$-like metric, but normalized so that a well-behaved measurement of a single star will have $RUWE \simeq 1$, regardless of primary color or magnitude. 
A number of studies have used a high RUWE as an indicator of binarity \cite[e.g.]{ziegler_soar_2019, pearce_orbital_2020}, and \citet{ziegler_soar_2019} found that $86\%$ of stars with $RUWE > 1.4$ had companions undetected by \gaia{} but seen in high-resolution follow up. The impact of an unseen companion on the astrometric fit (and hence RUWE) will depend on the separation and contrast of the companion. A sufficiently close companion will have a negligible impact on the observed PSF and will not generate a detectable astrometric wobble in the primary. A faint companion may impact the astrometric wobble, but not the PSF. 

\begin{figure*}[ht!] 
    \centering
    \includegraphics[width=\linewidth]{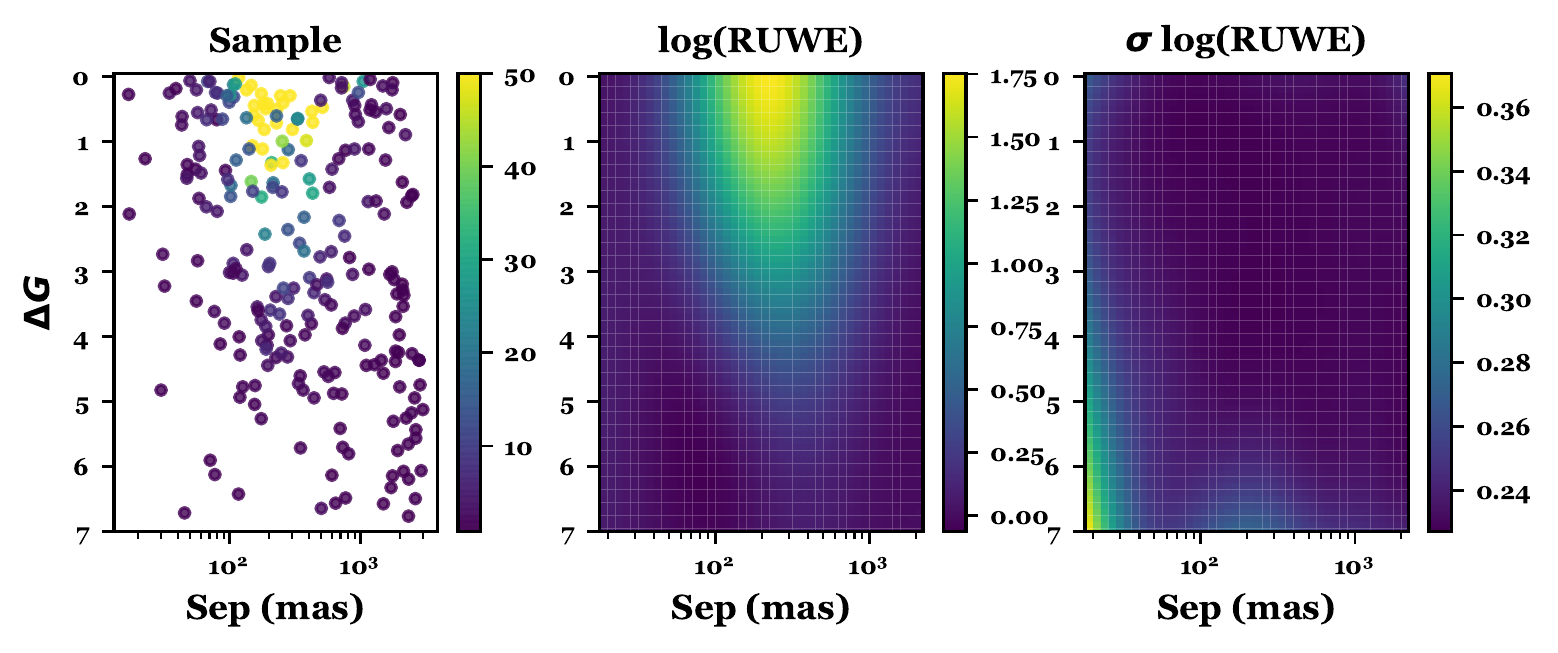}
    \caption{The sample of stars and fitted distribution for RUWE as a function of $\Delta G$ and projected separation. The first panel shows the input sample, color-coded by RUWE, the second the predicted $\log_{10}{(RUWE)}$ at each point of separation and contrast resulting from the Gaussian Process fit, and the third the predicted $\sigma_{\log{(RUWE)}}$.}
    \label{fig:RUWE_distribution}
\end{figure*}

To account for the relation between the contrast ($\Delta G$) and separation ($\rho$) of an unseen companion and the \gaia-determined RUWE, we use the  empirical calibration from Kraus et al. (in prep). By using an empirical calibration we can describe the behavior of \gaia{} RUWE in the presence of a companion, without a full description of the underlying cause (e.g., PSF mismatch or astrometric motion), or building a full model of the complex \gaia{} reduction. The calibration is based on a set of observations drawn primarily from Keck adaptive optics and non-redundant aperture masking as described in \citet{kraus_impact_2016}. Kraus et al. (in prep) includes additional data (with identical observational and reduction strategies), a description of how $\Delta G$ values are assigned depending on if the companion is bound or background, and the model fit to the RUWE data. To briefly summarize, $\Delta G$ is adopted from the \gaia{} catalog when the two are resolved, extrapolated from the $\Delta K$ detections in \citet{kraus_impact_2016}, or extrapolated from optical contrasts in the literature using the subset of companions resolved in \gaia. This output is fit using a Gaussian Process Kernel. The resulting model gives a prediction for $\log~$RUWE and $\sigma_{\log \rm{RUWE}}$ based on $\rho$ and $\Delta G$. The sample of known binaries and the resulting RUWE prediction model are shown in Figure \ref{fig:RUWE_distribution}. This shows that the RUWE of a binary star system is highest at separations between $100 - 1,000$ mas, and with contrast $<3$, decreasing at larger contrasts, and both larger and smaller separations. 

For every companion generated by \code, we calculate $\rho$ and $\Delta G$ at the average \gaia{} eDR3 epoch (2016.0). We use the model above to convert this to a predicted RUWE. \code{} then compares the observed eDR3 RUWE to the model-generated value using a one-sided Gaussian (half-normal) distribution to derive a rejection probability. We use a one-sided Gaussian distribution so that stars where the fit is unusually good (an unusually low RUWE) are not rejected, as there is no evidence this is indicative of binarity. This test is only applied within the range of stars over which the calibration is valid (approximately 20-2100\,mas and $\Delta G<7$). 
 
\begin{figure}[ht!] 
    \centering
    \includegraphics[width=\linewidth]{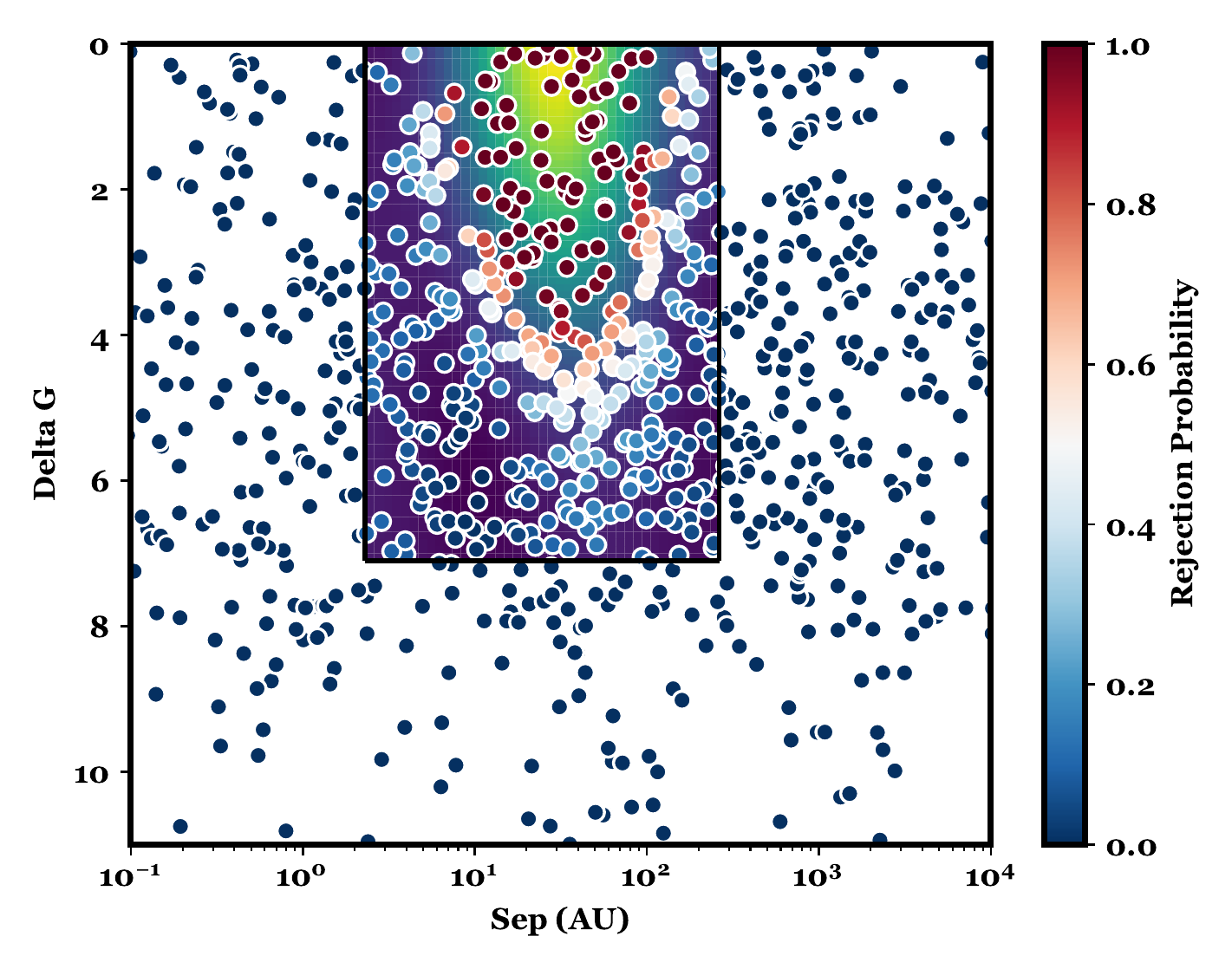}
    \caption{Figure showing the results of the RUWE test on a generated sample of 1,000 companions to HIP\,67522. Points represent the generated companions, with their projected separation and $\Delta G$ calculated as described in the text at an epoch of 2016.0. The points are colored according to their rejection probability, with  dark red indicating a higher rejection  probability, and dark blue representing a low rejection probability. The box and shaded region denote the separations and contrast at which the test applies, shaded according to the model-predicted $\log_{10}(RUWE)$. Outside of this box the test is not applied and no companions are rejected.}
    \label{fig:RUWE_results}
\end{figure}

Figure \ref{fig:RUWE_results} shows the results of applying just the RUWE comparison on a system with a low RUWE of 0.87 (HIP67522; see Section \ref{sec:hip67522} for details). In such a system, companions between $10$ and $50$ AU, with $\Delta G < \sim 4$ can be rejected nearly $100\%$ of the time, but this rapidly drops off with increasing contrast and changing separation.

\subsubsection{Additional Input}

There are less common situations or observational data that can provide separate constraints on the range of possible binaries around a given star. Some of these, like astrometric observations could be added by expanding the existing framework (e.g., the radial velocity and RUWE code can be expanded to include a model of the astrometric position for a given binary). Less common examples, like color limits on diluting companions from the chromaticity of transit or eclipse depths, can be applied to the output of \code{} after generation \citep[as was done in ][]{mann_tess_2020}. 

A somewhat common example that can be used in the existing code include limits from color-magnitude diagram (CMD) position. An unresolved binary should sit higher than a single star (all other parameters being equal) so a low position on a CMD can often be used to rule out nearly equal-mass companions. Such information can be included as a contrast curve as described in Section~\ref{sec:AO}, with no inner working limit and a flat contrast curve out to the size of the extraction aperture. The utility of such cases depends on knowledge of the metallicity and/or age of the star, which makes it hard to build in a general framework. Instead, we assume the user can provide the relevant contrast curve.  

\section{Test Cases} \label{sec:tests}

In the section we demonstrate \code{} on data from a number of different stars to showcase its effectiveness in combining multiple datasets and ruling out possible stellar companions. 
Some of the uses of \code{} that we highlight in this section include: ruling out false positives from stellar companions in planet detections; using \code{} on a star with a known stellar companion; using high-resolution imaging spanning a period of years to rule out companions; and estimating the number of missing stellar companions in a sample of stars.

\subsection{HIP67522} \label{sec:hip67522}

HIP~67522 is a 10-20 Myr old star in the Scorpius-Centaurus OB Association. \citet{rizzuto_tess_2020} detected a planet around it using data from \textit{TESS}, and followed up with a suite of ground-based facilities to rule out false positives, including RV monitoring and HRI. An early version of \code{} was used as part of the false positive analysis in that paper. Here we describe an updated analysis, where \code{} was used to combine the RV, HRI, and \gaia{} data to place stronger limits on the mass of any unseen companion.

\begin{figure*}[htb!]
    \begin{centering}
    \subfloat{\label{fig:hip67522_comp}\includegraphics[width=0.5\linewidth]{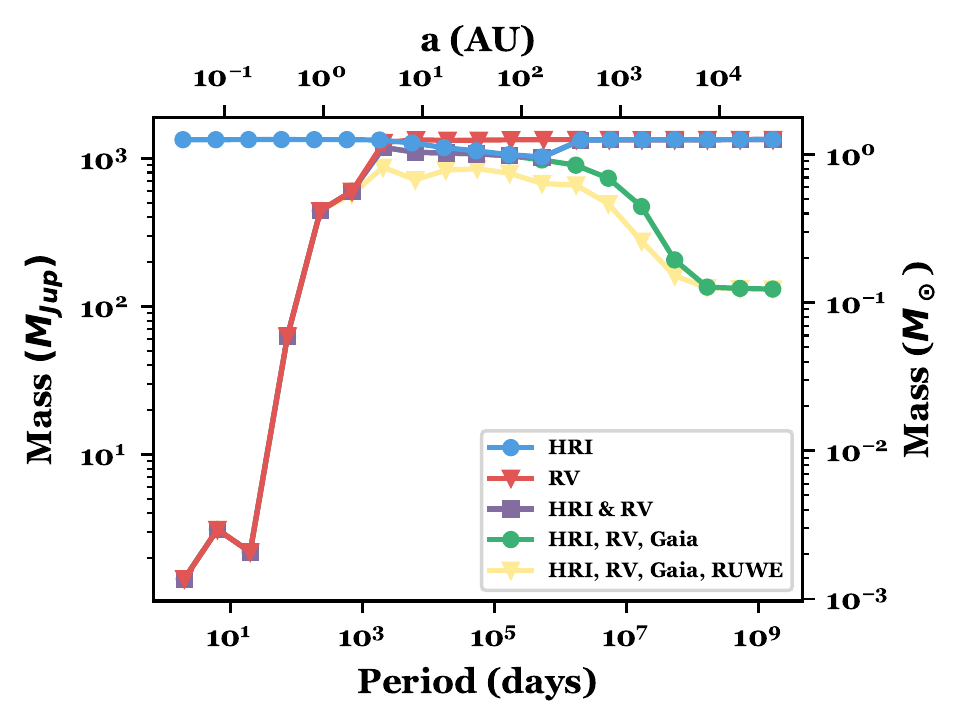}}
    \subfloat{\label{fig:hip67522_arg}\includegraphics[width=0.5\linewidth]{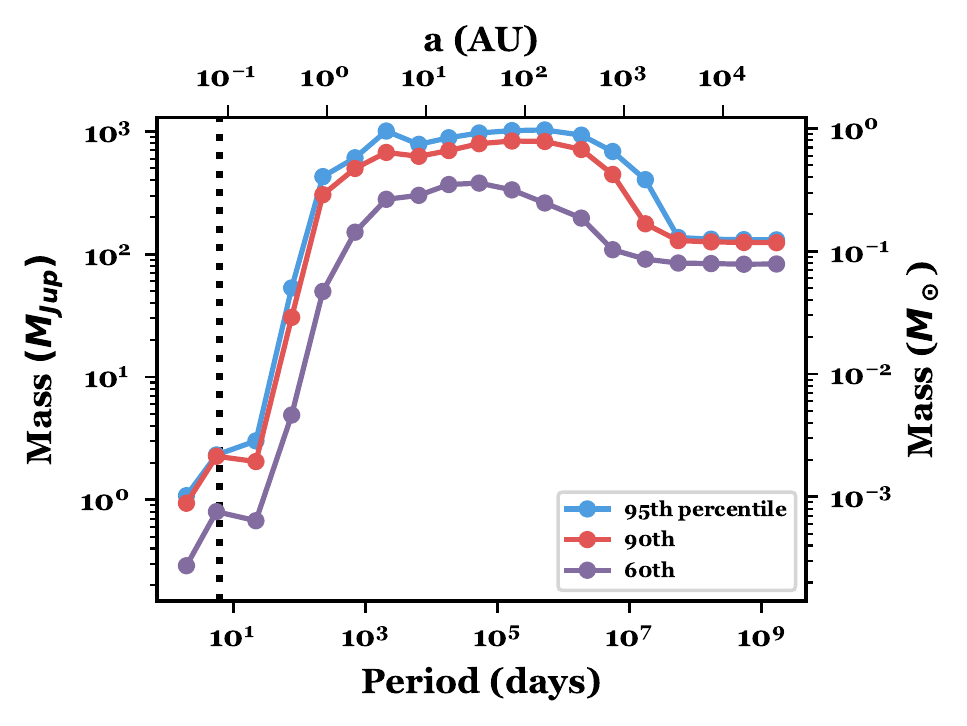}}\\
    \end{centering}
    \caption{Detection limits on the system HIP~67522:
    Left) 3$\sigma$ detection limits for transiting companions around HIP~67522 using different combinations of data. The color shows which datasets go into each resulting curve; HRI  only (blue), RV only (red), HRI and RV (purple), HRI, RV and Gaia imaging (green), and HRI, RV, Gaia imaging and RUWE (yellow). Imaging data looks less effective because of the log-log scaling (i.e., the shallow imaging used here is not effective for planetary-mass objects at this age), but it effectively removes stellar-mass objects at moderate separation. The RV limits are especially strong here because the inclination was restricted to only eclipsing/transiting objects. Similar to HRI, RUWE works best in projected separation, but by adding it in a larger number of companions at moderate periods can be eliminated, increasing the sensitivity between $10^3$ and $10^7$ days.
    Right) $1$, $2$, and $3\sigma$ Detection Limits for transiting companions around HIP 67522, using constraints from HRI, RV, \gaia{}imaging and RUWE. The detected planet at P$\sim$ 6 days has a $3\sigma$ mass detection limit of $ \sim 3 M_{Jup}$. Below $10^2$ days the RV measurements apply stringent limits on the mass of a companion, partially due to the locking to only transiting companions. When simulating non-transiting companions, the mass limit at $P \sim 6$ days is slightly higher at $\sim 8 M_{jup}$. Between $\sim 10^2$ and $\sim 10^5$ days neither the contrast nor the RV measurements can successfully rule out massive companions.}
    \label{fig:HIP67522_dtct_lims}
\end{figure*}

We simulate several sets of companions for HIP~67522 and analyze them with different combinations of the available data set, (e.g. just HRI, just RV, HRI and RV etc.) to demonstrate the utility of combining all possible data. Figure \ref{fig:hip67522_comp} shows the results of these tests. We apply an added jitter of 100\mps\ and a RV sensitivity floor of 20\mps\ for all tests based on the stellar variability of the star and limits of the RV instruments used.

To place constraints on the system we use the results from our simulation using the full combination of the data set, i.e. HRI, RV, \gaia{} imaging, and RUWE. RV measurements of HIP~67522 are taken from \citet{rizzuto_tess_2020}. An additional four RV measurements were taken on LCO/SALT between the publication of this paper and \citet{rizzuto_tess_2020}, and are presented in Table~\ref{table:HIP67522_RV}.
We simulate 5 million transiting companions and use them to examine the possibility of a false positive due to an eclipsing stellar companion by forcing the inclination of all companions to be consistent with an eclipsing system. We also simulate a second set of 5 million companions at all inclinations, which we use to explore the possibility of a stellar companion in the system at any inclination, which could alter the derived parameters.

\begin{table}[t]
    \centering
        \begin{tabular}{|c|c|c|c|c|c|} 
         \hline
         BJD  & RV & RVerr  & Telescope \\ 
         (days) & (\kms) & (\kms) & \\
         \hline
         \hline
        2458705.2675 & 7.95 & 0.33 & SALT \\
        2458709.2372 & 7.79 & 0.13 & SALT \\
        2458707.5159 & 8.52 & 0.85 & LCO \\
        2458708.4822 & 8.41 & 0.42 & LCO \\
         \hline
        \end{tabular}
    \caption{HIP67522 RV Measurements are taken from \citet{rizzuto_tess_2020} Table 4. An additional 4 RV measurements are taken on LCO/SALT between the publication of the paper and \citet{rizzuto_tess_2020}. These data are presented in this table.}
    \label{table:HIP67522_RV}
\end{table}

\begin{figure}[ht!] 
\includegraphics[width=\linewidth]{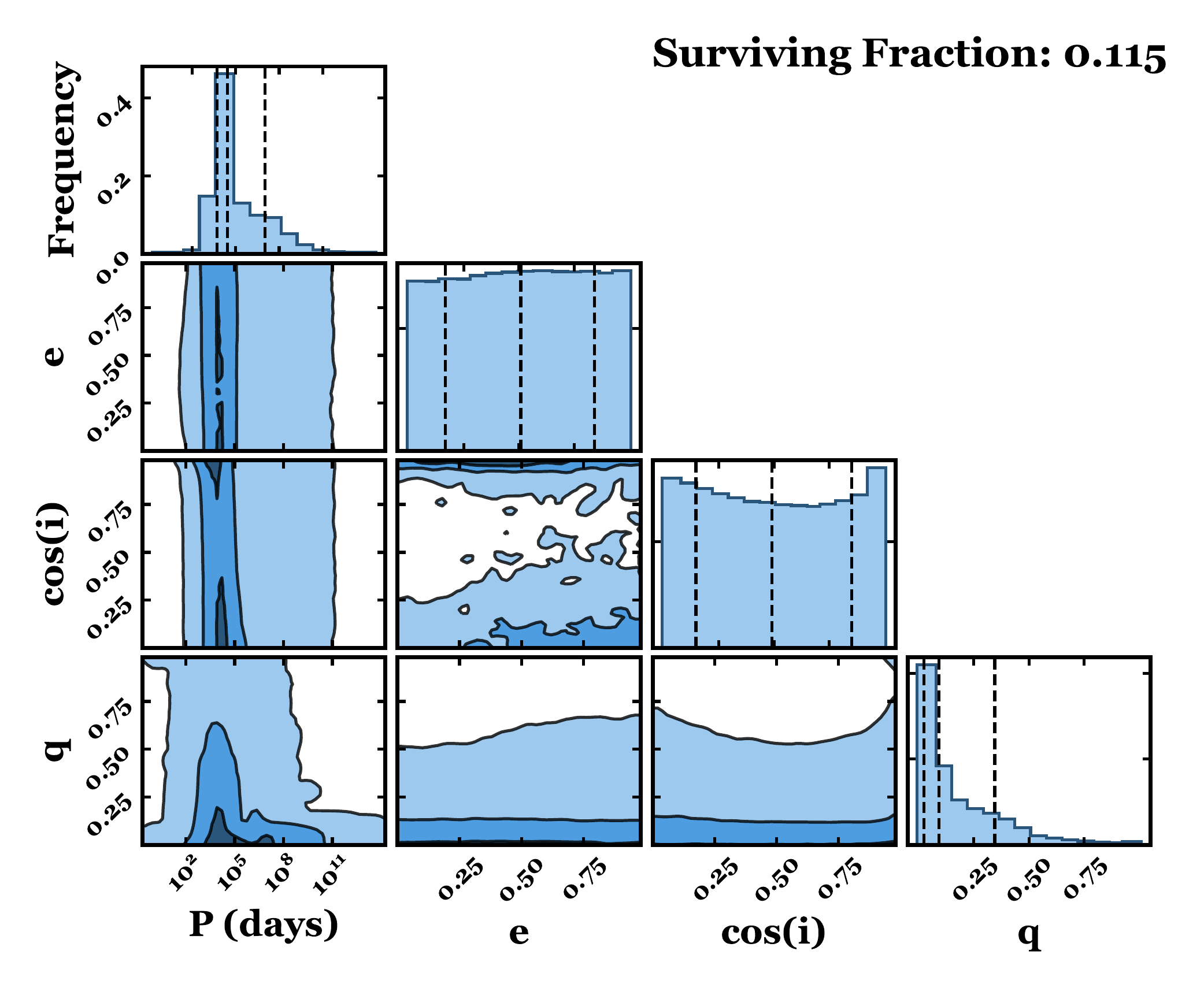}
\caption{Corner plot of major orbital parameters for surviving companions of a test combining HRI, RV, \gaia{} imaging and  RUWE limits for HIP~67522, with companions generated at all inclinations. Contours mark the 39.8, 86, and 98th percentiles of density. The top plots of each column show a histogram of that column's parameters. The dashed lines on the histograms mark the median and $\pm1\sigma$.}
\label{fig:hip67522_corner}
\end{figure}

 Results of the tests to compare different combinations of data are shown in Figure \ref{fig:hip67522_comp}. The combination of these four complementary tests provides powerful limits on companions. Radial velocity data, typically taken for exoplanet vetting, usually only covers close-in companions, relatively orthogonal to the other datasets. \gaia{} imaging is useful for wide binaries, and HRI and RUWE bridge the gap between the other two. The \gaia{} imaging and RUWE tests, operating at long and mid range periods, are able to rule out $\sim 50\%$ of binaries without the need for any additional follow-up data.

The effective range of HRI coincides with the effective ranges of RUWE and \gaia{} imaging in period (see Figure~\ref{fig:comparison}). HRI and \gaia{} imaging both work in projected separation, and overlap in that regime as well. Since HRI works best in projected separation, rather than period, the addition of HRI follow-up is unable to add significantly to mass detection limits, shown in Figure~\ref{fig:hip67522_comp}. However, the inclusion of HRI imaging can help to rule out a large number of companions overall, rejecting $17.1\%$ more companions than when not included.

Figure \ref{fig:HIP67522_dtct_lims} shows the mass detection limits produced by the search for transiting stellar-mass companions, such as could have caused a false planet detection. At the period of the detected planet we find a $3\sigma$ mass limit of $\sim 3 M_{Jup}$. This rules out the possibility of a false positive from a stellar mass companion orbiting at that period. Our limits are consistent with, but $\sim 2 M_{Jup}$ deeper than the findings presented in \citet{rizzuto_tess_2020}.

Figure \ref{fig:hip67522_corner} shows the results from the search for a companion at any inclination. We are able to rule out $88.4\%$ of the simulated companions. Surviving companions occur largely at low mass ratios. Short period ($P < 100$ days) companions are almost entirely ruled out, as are companions with high masses ($q>0.5$).

We are unable to definitively rule out a solar-mass companion at moderate periods ($10^3 < P < 10^7$ days). The HRI, RUWE and \gaia{} imaging tests are all effective in this period range, eliminating $81.9\%$ of the generated companions. However, among the small number of survivors are a non-trivial number of high-mass stars. Specifically, some stars happened to have projected separations close to the primary at the epoch of the observations. Such missed stars are rare; at short periods ($P < 10^3$ days) high mass companions are rejected by the radial velocity data regardless of projected separation, while at long periods ($P > 10^7$ days), companions spend a small fraction of their orbit behind the primary. Figure \ref{fig:HIP67522_dtct_lims} makes such situations appear common, because it was based on the mass distribution of \textit{surviving} companions. Effectively, this only indicates if a companion survives, what kind of companion parameters are possible. Depending on the science goal, this result can instead be interpreted alongside the overall survivor fraction or divided by the input distribution.

\subsection{DS Tuc A} \label{sec:ds_tuc}

DS Tuc A is a young star in a known binary system. \citet{newton_tess_2019} detected a transiting planet using observations from the \textit{TESS} survey, followed up with spectral observations from several ground based telescopes as part of the THYME young planets project.
The primary star, DS Tuc A, is a Sun-like star of spectral type G6V. The secondary star is at an angular separation of 5\arcsec, or $\sim6\times10^5$ days, with a spectral type K3V, and an estimated mass of $0.84\pm0.06 M_\odot$ \citep{newton_tess_2019}. 
The planet has a measured radius of $5.70\pm 0.17 R_{\oplus}$, and orbits with a period of $8.1$ days \citep{newton_tess_2019}.

With a separation of 5\arcsec and a contrast of $
\sim 1$ mag, DS Tuc AB is resolved in \gaia{}. Using the \gaia{} contrast curves constructed from \citet{arenou_gaia_2018, ziegler_measuring_2018, brandeker_contrast_2019} (see Section \ref{sec:Gaia} for details), we see that $\sim 100\%$ of companions in that parameter space are rejected. This means that, if applied, \gaia{} imaging will falsely rule out the known companion. This occurs because a core assumption of \code{} is that there are no detected companions. Since \code{} is designed to place constraints on \textit{undetected} companions, any companions that could have been detected, like DS Tuc B, are rejected. For this reason, it is important to not use datasets in which the effects of a known companion are present. In this example, we resolve the problem by leaving out the \gaia{} imaging data, and using only the RV measurements and \gaia{} astrometry. Since the known companion's period is much longer than the observational baseline of the RV data used, the known companion cannot be detected in it. Therefore, \code{} can use the RV data to probe for companions at shorter periods. Similarly, since the two companions are resolved in \gaia{}, the RUWE of the primary will not be elevated due to the known secondary, but could be elevated from a closer, unknown companion. We leave out the \gaia{} imaging altogether, but an alternative method would be to use a modified contrast curve for \gaia, excluding the region around the known companion, or a contrast curve which extends only as far as the nearest neighbor.

We simulate two sets of 5 million hypothetical companions, one with inclinations locked to only eclipsing companions, and one with all inclinations.
The simulated companions are compared to RV follow-up measurements, taken from  \citet{newton_tess_2019}, and RUWE. For the RV test, we used an added jitter of 100\,\mps\ and an RV sensitivity floor of 20\mps\ (again based on the expected stellar variability and instrument limits).

\begin{figure}[ht!] 
\includegraphics[width=\linewidth]{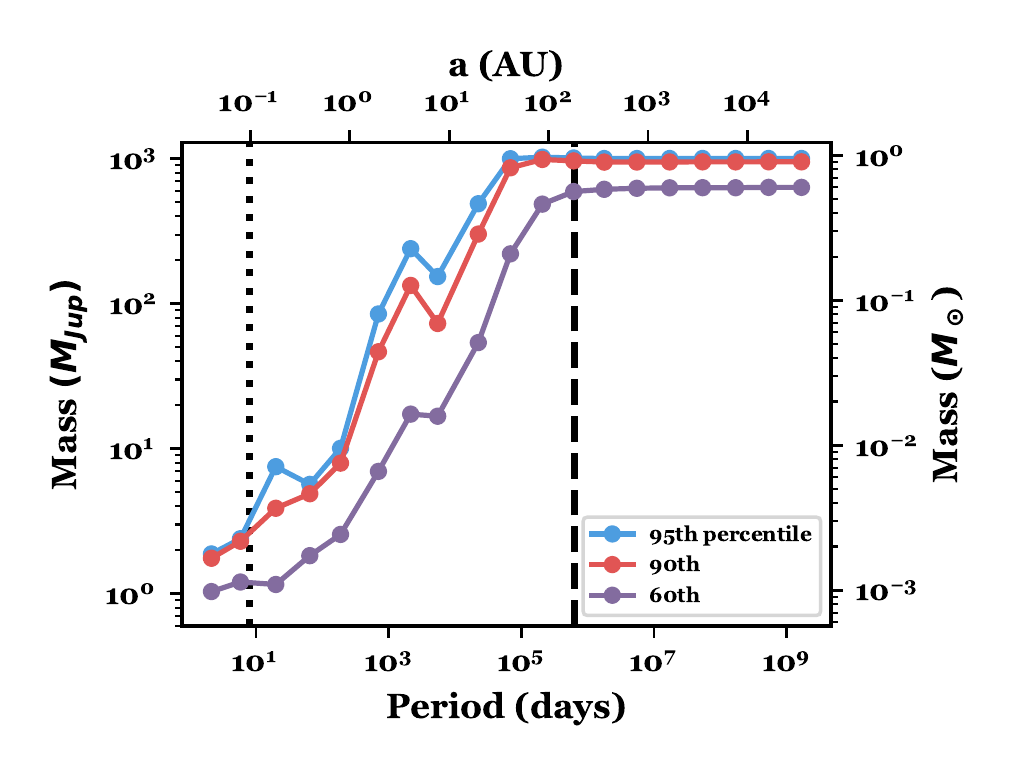} 
\caption{Detection Limits for transiting companions around DS Tuc A, using RV and RUWE constraints. The period of the detected planet is marked with a dotted line, and the period of the known stellar companion with a dashed line. The $3\sigma$ limit at the period of the planet is $\sim3 M_{Jup}$. At the estimated period of the known stellar companions the $3\sigma$ mass limit is $\sim 1 M_\odot$.}
\label{fig:ds_tuc_dtc_lim}
\end{figure}

\begin{figure}[ht!] 
\includegraphics[width=\linewidth]{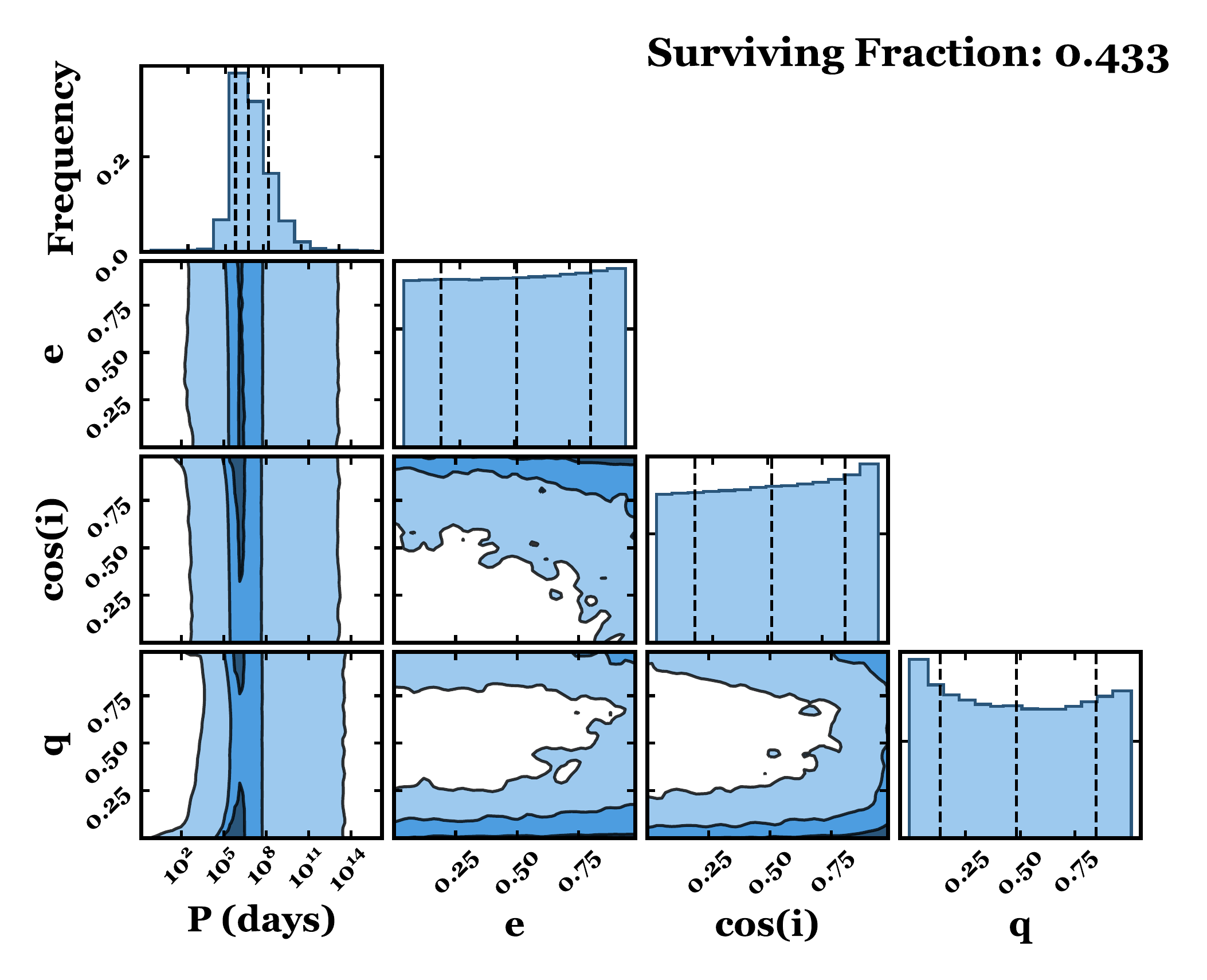}
\caption{Distributions of major orbital parameters for surviving companions around DS Tuc A, with all inclinations simulated.Contours mark the 39.8, 86, and 98th percentiles of density. The top plots of each column show a histogram of that column's parameters. The dashed lines on the histograms mark the median and $\pm1\sigma$.}
\label{fig:ds_tuc_corner}
\end{figure}

To determine the likelihood of a false positive detection from a stellar companion, we use the results from the simulation of only transiting companions. The mass detection limits for this simulation are shown in Figure~\ref{fig:ds_tuc_dtc_lim}. We find a $3 \sigma$ maximum mass of the detected companion at 8 days of $ \sim 3 M_{Jup}$, which rules out a stellar false positive.
At the period of the known stellar companion the maximum surviving mass was $\approx1M_\odot$. 

Similarly to the results in Section~\ref{sec:hip67522}, we are unable to rule out a stellar mass companion at moderate periods, here between $10^5$ and $10^9$ days. This is because there are a few generated companions at these periods with high $q$ but low $\rho$. At low $\rho$, they cannot be rejected by the RUWE test, which operates at longer projected separations, or the RV data, which operates at shorter periods. Since many of the companions at these periods are rejected by the RUWE test, the rare close-in, high-mass companions are numerous enough to increase the detection limits.

Figure~\ref{fig:ds_tuc_corner} shows the posterior distributions of the major parameters from the search for a companion at any inclination. We are able to rule out $56.7\%$ of the simulated companions. Surviving companions occur largely at low mass ratios and edge-on inclinations. Short period ($P<100$ days), and long period systems ($P>10^9$ days) are almost entirely ruled out, with the exception of very-low mass companions ($M < 0.1 M_\odot$). 
Very few high mass companions survive, and those that do are at high eccentricities and edge-on inclinations. Since it is possible that high eccentricity systems are more rare than reflected in our parameter generation (see Section \ref{sec:eccentricity} for details), the true limits may eliminate more stars than limits found by \code. 

\subsection{The ZEIT Sample} \label{sec:ZEIT}

In principle, \code{} can be run on large samples of stars and the likelihoods combined to set broad constraints on the overall binarity rate in the sample. This is important for work on the role of binarity on exoplanet occurrence rates, but can be similarly used to explore the role of binarity in any astrophysical situation \citep[e.g., clusters and high-mass stars,][]{gullikson_searching_2013, douglas_k2_2019}.
In practice, implementation depends on the sample, as the underlying prior of being a binary changes (e.g., exoplanet hosts are known to host fewer binaries than non-hosts \citep{wang_influence_2014-1, kraus_impact_2016}). 
While some studies on this topic forward model their data, e.g., taking a sample of binaries and attempting to reproduce the observed population by removing companions that would have been detected \citep[e.g.,][]{kraus_impact_2016}, others assume that any stellar companions present in their sample are detected in the data \citep[e.g.,][]{wang_influence_2014-1, wang_influence_2015, moe_impact_2019}. Many companions will be missed in such an analysis, especially over a large sample of targets. Instead, we can use \code{} to determine what kinds of possible companions could have survived on a star-by-star basis. As an example, we apply \code{} to the ZEIT (Zodiacal Exoplanets in Time) survey of young planets \citep{mann_zodiacal_2016-1}, and determine a completeness correction for binarity within the sample. We then compare the resulting corrected binary rate to that of a similar field population to see if differences in the companion population can help explain differences in the planet populations (as opposed to evolutionary changes).

The ZEIT survey found over a dozen planets in young clusters, most of which are given in \citet{rizzuto_zodiacal_2017}. We used a subset of 12 planet hosts from this survey to calculate a likely number of missing binaries in the group. One star from the sample, K2-136, is a known binary system, but all others are presumed to be single in the discovery papers due to the lack of detections. 

Many of the stars from the sample had either RV or HRI data, or both, summarized in Table \ref{table:ZEIT}. We use \code{} on each of the stars with any available RV or HRI data, and add constraints from \gaia{} imaging and RUWE. For all stars, we use a conservative jitter of 20\mps and an RV sensitivity floor of $20$\mps. We generated companions at all inclinations.
For each of the stars in the sample we determine the fraction of simulated companions which could not be rejected by the given data, listed in Table \ref{table:ZEIT}. Below, we highlight one of the stars which had a unique dataset that emphasizes the usefulness of \code{}.

\begin{table*}[ht!]
    \begin{centering}
        \begin{tabular}{|c|c|c|c|c|} 
         \hline
         K2 Name & Other Name & RV Ref. & Contrast Ref. & Survivor Fraction \\ 
         \hline
         \hline
        K2-25 & EPIC 210490365 & 1 & 1 & 0.265 \\
         \hline
        K2-33 & EPIC 205117205 & 3 & 3* & 0.426\\
         \hline
        K2-77 & EPIC 210363145 & & 2 & 0.513 \\
         \hline
        K2-95 & EPIC 211916756 & & 10 & 0.535 \\
         \hline
        K2-100 & EPIC 211990866 & 12 & 4 & 0.127 \\
         \hline
        K2-101 & EPIC 211913977 & & 4 & 0.441 \\
         \hline
        K2-102 & EPIC 211970147 & & 4 & 0.429 \\
         \hline
        K2-103 & EPIC 211822797 & & 4 & 0.485 \\
         \hline
        K2-136 & EPIC 247589423 & 6 & 9 & -- \\
         \hline
        K2-264 & EPIC 211964830 & & 11 & 0.605 \\
         \hline
        - & EPIC 211901114 & & 4\textdagger{} & 0.627 \\
         \hline
        - & HD 283869 & & & 0.493 \\
         \hline
        \end{tabular}
    \caption{The ZEIT Sample; *K2-33 contrasts included one additional contrast curve, taken in 2011 which was not included in the analysis of (3). \textdagger{} EPIC 211901114 contrast acquired after publication, but using the same observational setup as described in (4); References: (1)\cite{mann_zodiacal_2016} (2) \cite{gaidos_zodiacal_2017} (3) \cite{mann_zodiacal_2016-1} (4)\cite{mann_zodiacal_2017} (5) \cite{rizzuto_zodiacal_2017} (6) \cite{mann_zodiacal_2017-1} (7) \cite{vanderburg_zodiacal_2018} (8) \cite{rizzuto_zodiacal_2018} (9) \cite{ciardi_k2-136_2018} (10)\cite{obermeier_k2_2016} (11) \cite{livingston_k2-264_2019}  (12) \cite{barragan_radial_2019} }
    \label{table:ZEIT}
    \end{centering}
\end{table*}

\begin{center}
    \textit{K2-33}
\end{center}
K2-33 is an example of how multiple contrast curves can be combined for deeper limits than is possible with either individually. This is possible because \code{} calculates the projected separation of each companion before comparing them to the contrast curve, and the projected separations of the simulated companions may be different at two different observational epochs. For K2-33, we combine two HRI contrast curves, generated as described in \cite{mann_zodiacal_2016-1}. The first was taken the night of May 15, 2011, and the second the night of March 19, 2016, giving an observational baseline of nearly 5 years.

As a test case on the effectiveness of combining contrast curves, we ran \code{} using each contrast curve of K2-33 alone, and then using both contrast curves together. To showcase the maximum potential of this scenario, we set the distance to the star to 10pc, so that the HRI was more powerful. This distance is also more consistent to those of \tess{} stars, which could benefit from similar analysis. The survivor fraction listed in Table \ref{table:ZEIT} was calculated using both curves together, and the correct distance to K2-33.

Individually, each contrast curve could reject $\sim 25 \%$ of the simulated companions. Together they rejected $27.6\%$ of the simulated companions, an improvement of $2.6\%$. This is a modest improvement, but may be significant in other cases.
Similarly, one could combine multiple datasets using images taken at similar times but in different filters.

\begin{center}
    \textit{Binarity Rate of Planet Hosts}
\end{center}

We estimate the number of binary systems expected in a sample of field stars using a M-dwarf multiplicity rate of $26.8\pm1.4\%$ from \citet{winters_solar_2019}, and a G-star multiplicity rate of $46\pm2\%$ from \citet{raghavan_survey_2010}. The ZEIT sample contains one G-type, five M-type, and six K-type stars. We applied the G-star multiplicity rate to the G-star and the three early K-type stars, and the M-star multiplicity rate to the M-dwarfs and the three late K-type stars. From this we estimate that a similar sample of field stars would contain $4.0\pm 0.23$ binary systems. The binomial probability of detecting 1 or fewer companions is small (1.3\%). Although not statistically significant, it highlights the need to apply a correction for the unseen systems.

To estimate the completeness correction for binarity in the ZEIT sample, we sum the surviving fractions of each star, and multiply by the appropriate multiplicity rate, giving a correction of $1.61$. By adding $1$ to account for the known binary K2-136AB we calculate a predicted number of binaries of $\simeq2.61$, marginally lower than the number expected assuming a field-like population. The binomial probability of detecting $3$ or fewer binaries in a sample of 12, given an expected value of $4.0$, is $39\%$. The results are similarly consistent with the lower rate of binaries seen in older planetary systems, indicating that the sample used here is too small for robust results.

While the sample size that we use here is too small to draw conclusions, we highlight that the number of `unseeen' companions added in from our \code{} tests is larger than the number of detected systems, highlighting the need for such a correction. Further,  \code{} could be used to calculate binarity completeness corrections in any stellar population, facilitating studies comparing binarity rates between different populations. Here we simply added the resulting binary probabilities together to predict a combined binarity correction, but a more robust way to combine the results would be to combine likelihoods computed from \code{} using Hierarchical Bayes \cite[e.g.][]{hogg_inferring_2010}, allowing additional understanding of both the number and distribution (e.g., in separation and mass ratio) of companions in the sample of interest.

\section{Summary and Conclusions} \label{sec:conclusions}

We present here a framework and code, \code, which uses complementary types of stellar observations to place the tightest constraints on possible unseen stellar companions. In addition to the extra constraints of simultaneously fitting multiple datasets, the code properly accounts for differences between projected separation (what high-resolution imaging measures) and true semi-major axis (which if not done properly overestimates the advantages of such data), and takes advantage of \gaia{} imaging and RUWE to rule out a large set of companions even without additional follow-up data. 

The default parameters are for solar-type stars, given our choice of models with a maximum mass of $1.4 M_\odot$, and that the default prior distributions used for orbital parameters apply only for stellar companions, and not for planetary ones. The user can adjust this  as desired by changing to a different model or changing the prior distributions.

We test this code on two young stellar systems with detected planetary companions to place limits on the masses of the companions, and on a sample of young planets without known companions to calculate a likely number of undetected stellar companions.

We use the code on the star HIP~67522, and place a mass limit on a transiting companion at the detected period of $\sim 3 M_{Jup}$, ruling out a stellar or brown dwarf mass object as the source of the signal. Examining all possible companions, we eliminate $88.5\%$ of simulated binaries.
For the known binary DS~Tuc\,AB, we place a $3\sigma$ mass limit at the period of the newly-detected companion, of $\sim 3M_{jup}$, ruling out a stellar false positive. In addition, $ 56.7\% $  of companions generated at all inclinations are ruled out.

By running \code{} on a sample of stars from the ZEIT survey we calculate a likely number of $1.6$ undetected companions  for that sample, and determine that in this small sample, the binarity rate is lower, but consistent with field rates. This small sample is not large enough to draw any conclusions, and a better analysis would require a sample of hundreds of stars, rather than the dozen that we analyzed for demonstration purposes.

Even with full follow up data (RV and HRI measurements), we are often unable to completely rule out a stellar-mass companion, as discussed in \ref{sec:hip67522}. This is because at moderate periods a non-trivial number of massive stars are on the inner parts of their orbits, where they cannot be ruled out by HRI, and where RV is ineffective. Even though this is a rare event, it is common enough to have a notable effect on the surviving companion statistics, since most of the simulated companions are eliminated.

\code{} could also be used for prioritizing observational follow-up. If only constraints from \gaia{} are sufficient for false-positive analysis, they may make additional HRI observations unnecessary or help optimize which HRI setup/instrument to use. As another example, the code could be used on different simulated sets of RV measurements to determine the frequency and precision of RV results required.

Future versions of this code could incorporate additional information from future data releases from \gaia{}, such as individual measurements of astrometry, or multi-epoch RV measurements. In addition to being more accurate than our use of RUWE, direct fitting of astrometry should significantly improve sensitivity to binaries with separations or inclinations that are difficult to detect with RV data but are too tight to detect with most HRI. Multi-epoch RV from \gaia{} could replace the need for additional ground-based follow-up.
With these we would be able to effectively search for binary companions in an even larger parameter space, without any follow-up observations. 

\section{Acknowledgements}
We thank the referee for their useful feedback and comments, which improved the paper. We also thank  Aaron Rizzuto for providing the additional RV measurements for HIP67522, and Greg Sloan for providing feedback on the manuscript. AWM and MLW were supported by NASA grant 80NSSC19K0583, MLW was also supported through NASA's \ktwo\ GO program (80NSSC19K0097) .

This work has made use of data from the European Space Agency (ESA) mission \emph{Gaia} \footnote{\url{https://www.cosmos.esa.int/gaia}}, processed by the \emph{Gaia} Data Processing and Analysis Consortium (DPAC)\footnote{\url{https://www.cosmos.esa.int/web/gaia/dpac/consortium}}. Funding for the DPAC has been provided by national institutions, in particular the institutions participating in the \emph{Gaia} Multilateral Agreement

This research has made use of the Keck Observatory Archive (KOA), which is operated by the W. M. Keck Observatory and the NASA Exoplanet Science Institute (NExScI), under contract with the National Aeronautics and Space Administration. This research has made use of the Exoplanet Follow-up Observation  Program  website, which  is operated by the California Institute of Technology, under contract with the National Aeronautics and Space Administration under the Exoplanet  Exploration Program.

\vspace{5mm}
\facilities{SALT (HRS), LCO (NRES)}

\software{Astropy \citep{astropy_2013, astropy_2018}, Astroquery \citep{ginsburg_astroquery_2019}, matplotlib \citep{matplotlib_2007},
NumPy \citep{numpy_2020},
SciPy \citep{scipy_2020}}

\clearpage
\nocite{*}
\bibliography{binary_bib}

\end{document}